\documentclass[apj]{emulateapj}

\usepackage{graphicx}
\usepackage{amssymb}
\usepackage{multirow}
\usepackage{float}
\usepackage{comment}

\newcommand{\psrj}{\hbox{\object{PSR\,J0108-1431}}}
\def\beq{\begin{equation}}   \def\eeq{\end{equation}}
\def\beqr{\begin{eqnarray}}   \def\eeqr{\end{eqnarray}}

\begin{document}
 
\title{XMM-Newton observation of the very old pulsar J0108-1431}
\author{B. Posselt}
\affil{Department of Astronomy \& Astrophysics, Pennsylvania State University, 525 Davey Lab,University Park, PA 16802, USA }
\email{posselt@psu.edu}
\author{P. Arumugasamy}
\affil{Department of Astronomy \& Astrophysics, Pennsylvania State University, 525 Davey Lab,University Park, PA 16802, USA}
\author{G. G. Pavlov}
\affil{Department of Astronomy \& Astrophysics, Pennsylvania State University, 525 Davey Lab,University Park, PA 16802, USA}
\affil{St.-Petersburg State Polytechnic University, Polytekhnicheskaya ul. 29 195251, Russia}
\author{R. N. Manchester}
\affil{CSIRO Astronomy and Space Science, Australia Telescope National Facility, P.O. Box 76, Epping NSW 1710, Australia}
\author{R. M. Shannon}
\affil{CSIRO Astronomy and Space Science, Australia Telescope National Facility, P.O. Box 76, Epping NSW 1710, Australia}
\author{O. Kargaltsev}
\affil{Department of Physics, The George Washington University, Washington, DC 20052, USA}

\begin{abstract}
We report on an X-ray observation of the 166\,Myr old radio pulsar J0108-1431 with XMM-$Newton$.
The X-ray spectrum can be described by a power-law model with a relatively steep photon index $\Gamma \approx 3$ or by a combination of thermal and non-thermal components, e.g., a power-law component with fixed photon index $\Gamma = 2$ plus a blackbody component with a temperature of $kT=0.11$\,keV. 
The two-component model appears more reasonable considering different estimates for the hydrogen column density $N_H$.
The non-thermal X-ray efficiency in the single power-law model is $\eta^{\rm PL}_{\rm 1-10\,\rm{keV}}= L^{\rm PL}_{1-10\,\rm{keV}}/\dot{E} \sim 0.003$, 
higher than in most other X-ray detected pulsars.
In the case of the combined model, the non-thermal and thermal X-ray efficiencies are
even higher, $\eta^{PL}_{\rm 1-10\,\rm{keV}} \sim \eta^{bb}_{\rm PC} \sim 0.006$.
We detected X-ray pulsations at the radio period of $P \approx 0.808$\,s with significance of $\approx 7\sigma$.  The pulse shape in the folded X-ray lightcurve (0.15--2\,keV) is asymmetric, with statistically significant contributions from up to 5 leading harmonics. Pulse profiles at two different energy ranges differ slightly: the profile is asymmetric at low energies, 0.15--1\,keV, while at higher energies, 1-2\,keV, it has a nearly sinusodial shape. The radio pulse peak leads the 0.15--2\,keV X-ray pulse peak by $\triangle \phi=0.06 \pm 0.03$.
\end{abstract}

\keywords{ pulsars: individual (PSR J0108--1431) ---
        stars: neutron ---
         X-rays: stars}

\section{Introduction}
Radiation of rotation powered pulsars (RPPs) is powered by their rotational energy loss. Over 1700 of RPPs have been detected in the radio and over 100 in X-rays. Their period and period derivative measurements are used to calculate the spin-down power, $\dot{E}$, the surface magnetic field $B$, and the characteristic spin-down age, $\tau$. The X-ray spectra of RPPs exhibit some common features that evolve with spin-down age.

For very young pulsars ($\tau \lesssim 10$\,kyr), the magnetospheric emission is observed to be dominant, burying the component of bulk surface thermal emission in most cases. Middle-aged pulsars, with spin-down ages $10^4 \lesssim \tau \lesssim 10^6$\,yr, tend to have a significant contribution from the surface thermal emission ($kT \sim 0.03 - 0.3$\,keV) in the soft X-rays, eventually dominating the X-ray spectrum up to $ \sim 1 - 2$\,keV. The high-energy part of the X-ray spectrum contains a contribution from non-thermal, magnetospheric emission and may contain a component of thermal, polar-cap emission. The non-thermal component in the spectra of these young pulsars is usually fit with a power-law (PL) with photon index $1 \lesssim \Gamma \lesssim 2$  (see, e.g., Figure 4 in \citealt{Li2008}). The fraction of the spin-down power emitted in the X-rays (the X-ray efficiency $\eta_{\rm X} = L_{\rm X}/\dot{E}$) has values in the range of $10^{-5} \lesssim \eta_{\rm X} \lesssim 10^{-3}$ (see e.g., Figure 5 in \citealt{Kargaltsev2008}). The X-ray pulse profiles of young pulsars show relatively sharp pulses for the non-thermal emission and smooth, broader pulses for the thermal emission. Asymmetric pulse shapes have been reported for middle-aged pulsars. For instance, Geminga, PSR\,B0656+14 and PSR\,B1055-52 ($\tau \sim$ a few 100\,kyr) and PSR J0538+2817 ($\tau = 30$ kyr) show such asymmetric pulse profiles (e.g., \citealt{deLuca2005,ZavlinP2004,McGowan2003,Pavlov2002}).

Old ($\tau \gtrsim 10^6$ yr) RPPs have diminished spin-down powers ($\dot{E} \lesssim 10^{34}$\,erg\,s$^{-1}$) and X-ray luminosities. This restricts the distance up to which such old pulsars can be detected and necessitates longer observations. There have been only a dozen of these old RPPs detected in the X-rays (see, e.g., \citealt{Posselt2012} and references therein). Absorbed PL model fits for these pulsars yield $\Gamma \sim 2 - 4$. Hence, their spectra are softer than the non-thermal spectra of the younger pulsar population. The inferred absorbing Hydrogen column density values, $N_H$, are usually larger than expected from the respective total Galactic Hydrogen column densities and/or estimates from pulsar dispersion measures. The non-thermal X-ray efficiencies in the 1--10\,keV band show significant scatter, $\eta_{1-10\;\rm{keV}} \sim 10^{-4} - 10^{-2}$, but are on average higher than the corresponding values calculated for younger pulsars \citep{Kargaltsev2006a, Zharikov2006}. \\

Old neutron stars are too cold to have significant thermal X-ray emission from the bulk stellar surface \citep{Yakovlev2004}. However, a possible thermal X-ray contribution may come from polar caps heated by infalling accelerated particles \citep{Harding2001,Harding2002}.
An additional thermal component can explain the X-ray spectra of some of these old pulsars. If fitted with a simple blackbody (BB) model, these components have BB temperatures of  0.1--0.3\,keV and projected areas smaller than the conventional polar-cap area ($A_{\rm{pc}} = 2\pi^2 R_{\rm{NS}}^3/(c P) \sim 10^5$\,m$^2$ for NS radius $R_{\rm{NS}} \sim 10$ km and period, $P \sim 1$\,s). Thermal emission from the surface layers of a neutron star atmosphere can differ substantially from BB emission (e.g., \citealt{Pavlov1995b}). 
Assuming H or He atmosphere emission instead of the simple BB results in larger polar-cap area sizes and temperatures about a factor 2 lower than the BB temperature
(see, e.g.,  \citealt{Zavlin2004b0950} for the atmosphere model analysis of the old PSR B0950+08).\\ 

Pulsar J0108--1431 was discovered by \cite{Tauris1994}. It has a period $P = 0.808$ s and a period derivative $\dot{P} = 7.7 \times 10^{-17}$ s s$^{-1}$, which corresponds to $\tau = P/2\dot{P} \approx 166$ Myr, $\dot{E} \approx 5.8 \times 10^{30}$ erg s$^{-1}$, and $B=2.5 \times 10^{11}$\,G, as listed in the Australian Telescope National Facility (ATNF) Pulsar Catalogue\footnote{http://www.atnf.csiro.au/research/pulsar/psrcat} \citep{Manchester2005}. 
PSR\,J0108$-$1431 has
been well monitored for two decades at radio wavelengths, and details on its
radio properties have been presented by, e.g, \citet{Espinoza2011} and \citet{Hobbs2010, Hobbs2004}. \citet{Weltevrede2008} discussed the high
linear polarization of $\sim 70$\,\% at 20\,cm, which is very unusual
for a pulsar with such a low $\dot{E}$ (see their
Figure\,8). Because of suspected similarities with high-$\dot{E}$ pulsars,
PSR\,J0108$-$1431 is now regularly monitored as part of the
collaborative program with the {\it Fermi} group
\citep{Weltevrede2010}. An optical counterpart of PSR J0108--1431 was suggested by \citet{Mignani2008}.\\

PSR\,J0108--1431 has the largest $\tau$ and lowest $\dot{E}$ among the X-ray detected pulsars. Characterization of its spectrum is essential to understand the final stages of the spectral evolution of old pulsars. It is one of the nearest neutron stars to Earth, with an estimated VLBI parallax distance of $d = 240^{+124}_{-61}$ pc \citep{Deller2009}, which was recently revised to $d = 210^{+90}_{-50}$\,pc when accounting for the Lutz-Kelkar bias \citep{Verbiest2012}. \citet{Deller2009} also measured the pulsar  proper motion of  $170.0 \pm 2.8$\,mas\,yr$^{-1}$ in the south-southeast direction.\\

PSR\,J0108--1431 has been detected with the {\em Chandra X-ray Observatory} \citep{Pavlov2009}. 
The spectral analysis, performed using the 51 counts detected in the 0.8--5\,keV energy range, showed emission consistent with either a single-component, soft ($\Gamma \sim 2.2 - 3.4$) PL expected from the pulsar's magnetosphere or a 3\,MK thermal BB emission from an apparent emitting area of  $\sim 50$ m$^2$.  \citet{Pavlov2009} suggested the presence of both non-thermal (magnetospheric) and thermal (polar-cap) components in the spectrum, potentially distinguishable from each other by phase-resolved spectroscopy of the source.
\begin{figure}[b]
\begin{center}
{\includegraphics[width=79mm,bb= 11 11 450 450,clip,height=80mm]{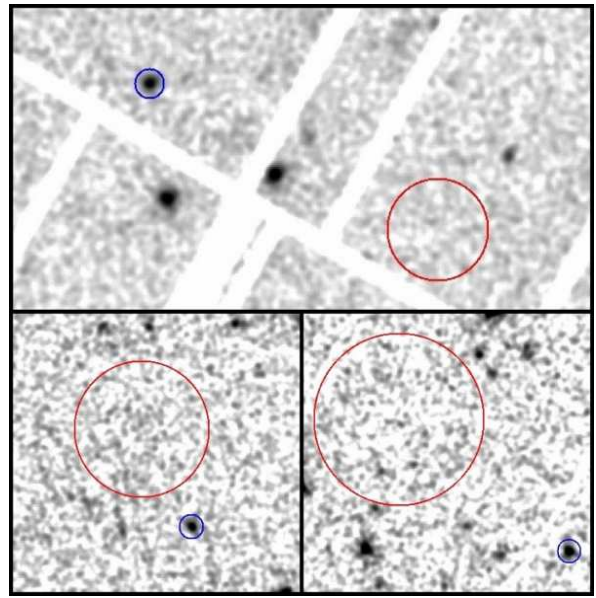}} \\
\caption{X-ray images (0.2-2.5\,keV) of the field in the direction of the PSR J0108$-$1431. The pn image ($\sim 4\arcmin \times 2\arcmin$),  and the MOS1 and MOS2 images (each $\sim 2.6\arcmin \times 2.6\arcmin$) are shown on the top, bottom-left and bottom-right, respectively. North is up, East is to the left. The source extraction regions (radii of $12^{\prime\prime}$ for pn, $13^{\prime\prime}$ for MOS) are marked with blue circles, the background regions are marked with red circles.
\label{regions}}
\end{center}
\end{figure}

The goal of our deeper {\em XMM-Newton} observation was to collect more photons for studying the pulsar spectrum. In addition, the high time resolution of the EPIC-pn detector could detect X-ray pulsations and potentially separate its non-thermal and thermal components through phase-resolved spectroscopy. The observation and data analysis are described in Section 2, the relation between the detected X-ray and radio pulse shapes are described in Section 3, followed by a discussion of the results in Section 4.

\section{Observation and Data Analysis}
\begin{figure}[t]
\begin{center}
\includegraphics[width=8.5cm,bb=40 14 575 402,clip]{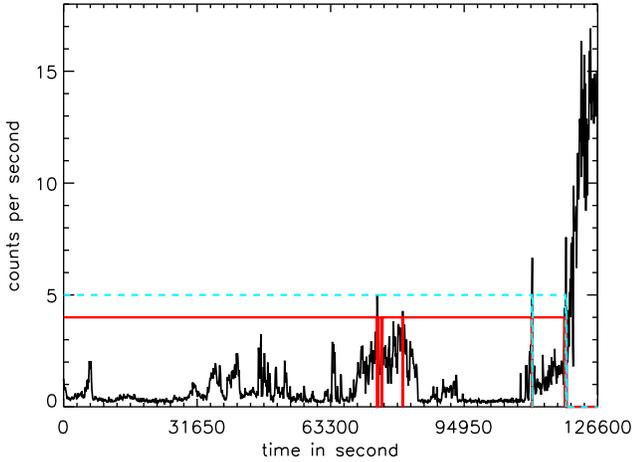}
\caption{Background light curve of the EPIC-pn detector in the energy range 10 to 12\,keV, binned to 100\,s.
The $x$-axis denotes the time counted from the obervation start, the $y$ axis denotes the count rate in counts\,s$^{-1}$.  
The red solid line shows the considered exposure time range for the chosen GTIs for our timing analysis, requiring a count rate $< 4$ counts\,s$^{-1}$. Similarly, the cyan line shows the chosen GTIs for our spectral analysis, requiring a count rate $< 5$ counts\,s$^{-1}$. Vertical lines indicate corresponding regions of flagged exposure times.   
\label{PNrate}}
\end{center}
\end{figure}

Pulsar J0108$-$1431 was observed on 2011 June 15 (MJD\,55727) with XMM{\em -Newton} (obsid 0670750101) using the European Photon Imaging Camera (EPIC) in Full-Frame mode with the Thin filter. 
The EPIC-pn \citep{Strueder2001} and EPIC-MOS1 and MOS2 cameras \citep{Turner2001} observed the source for 126.7 ks and 127.8 ks respectively.
The EPIC data processing was done with the XMM{\em-Newton} Science Analysis System (SAS) 11.0\footnote{http://xmm.esac.esa.int/sas}, applying standard tasks.
Our observation was contaminated by soft proton flaring events in the background.
We considered different selections of Good Time Intervals (GTIs) for optimal spectral and timing analysis, which are shown in Figure~\ref{PNrate} and described in more detail below.

We detected the X-ray counterpart of PSR\,J0108$-$1431 at a position $\alpha = 01^{\rm h} 08^{\rm m} 08\fs{25} , \delta = -14^\circ 31^\prime 53\farcs{4}$ with EPIC-pn.
This position is separated by $2\farcs{9}$ from the earlier (MJD\,54136) {\em Chandra} detection at  $\alpha = 01^{\rm h} 08^{\rm m} 08\fs{354} ,  \delta = -14^\circ 31^\prime 50\farcs{38}$ \citep{Pavlov2009} after accounting for the proper motion by \citet{Deller2009}.
The X-ray position differs by $3\farcs{0}$ from the expected radio pulsar position using the proper motion and VLBI position: $\alpha = 01^{\rm h} 08^{\rm m} 08\fs{347} ,  \delta = -14^\circ 31^\prime 50\farcs{187}$ (MJD\,54100) listed by \citet{Deller2009}.
The absolute astrometric $2\sigma$ error of  XMM{\em -Newton} is $4^{\prime\prime}$ \footnote{XMM calibration document: \protect\\ http://xmm2.esac.esa.int/docs/documents/CAL-TN-0018.pdf}. Thus, the EPIC-pn X-ray position is consistent with the {\em Chandra} and VLBI radio position.

\subsection{Spectral Analysis}
\label{spectralana}

For our spectral analysis of the X-ray counterpart of PSR J0108$-$1431, we aimed for the highest possible signal-to-noise ratio, S/N = $ (N_s - a N_b)/(N_s + a^2  N_b)^{\frac{1}{2}} $, where $N_s$ and $N_b$ are the total numbers of counts extracted from the source and background regions of areas $A_s$ and $A_b$, respectively, and  $a = A_s/A_b$.
Given the strong background flaring in our observation (Figure~\ref{PNrate}), we tested different GTI filters derived from 100\,s binned light-curves of events with energies above 10 and 12 keV for pn and MOS detectors, respectively.
For each GTI-filtered event file, we applied the \texttt{eregionanalyse} task to optimize the aperture size for high S/N spectral extraction. Standard pattern filtering ($\leq$ 4 for pn and $\leq$ 12 for MOS) was enforced.
We achieved the highest S/N for a GTI count rate cut-off of 5.0\,counts\,s$^{-1}$ and 0.5\, counts\,s$^{-1}$ for the pn and MOS light-curves, and for source aperture radii of $12^{\prime\prime}$ and $13^{\prime\prime}$ for pn and MOS, respectively.  
The net GTIs for pn, MOS1 and MOS2 are 104.8\,ks, 112.5\,ks, and 114.6\,ks, respectively.\\

\begin{figure}[t]
{\includegraphics[width=60mm,height=80mm, angle=270]{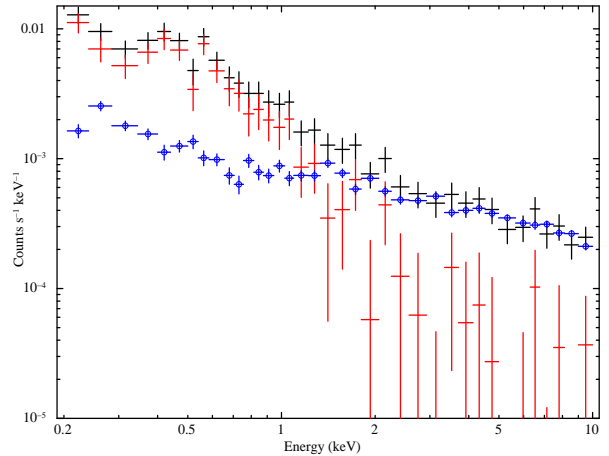}} \\
\caption{Count rate spectra of the binned counts (minimal S/N$\ge 5$) in the source region (black), background counts (blue circles), and  
background subtracted source counts (red) for EPIC-pn.
\label{background}}
\end{figure}

We used larger background regions for better statistics, with the radii of 42$^{\prime\prime}$, 75$^{\prime\prime}$ and 90$^{\prime\prime}$ for pn, MOS1 and MOS2, respectively. 
The extraction parameters are presented in Table \ref{specfilter}.
The source and background regions in the three processed EPIC images are shown in Figure~\ref{regions}. 
The background signal is comparable to the source signal at 1.2 to 2\,keV and then exceeds the source signal for energies $>3$\,keV (Figure~\ref{background}). 
Considering the substantial count rate errors for the high energies, we restrict our spectral fitting to events in the 0.2--2.5\,keV range in order to better constrain the spectral fit parameters.

Considering all three EPIC cameras together, we obtained around 990 total X-ray counts from the source apertures, or around 705 source counts (see Table~\ref{specfilter}).  Redistribution matrices and effective area files were generated using the usual SAS tasks \texttt{rmfgen} and \texttt{arfgen}. For the spectral fit, the SAS task \texttt{specgroup} was used to group the source counts of each spectra with a minimal S/N$\ge 5$ per bin.\\ 

\begin{table}[]
\caption{Spectral extraction parameters for pn, MOS 1/2 events.\label{specfilter}}
\begin{tabular}{lccc}
\hline
Extraction parameters	& pn		& MOS 1		& MOS 2		\\  \tableline
Aperture radius	&  $12^{\prime\prime}$	&  $13^{\prime\prime}$ &  $13^{\prime\prime}$\\ 
Energy range	(keV)	& 0.2 - 2.5	& 0.3 - 2.5	& 0.3 - 2.5	\\ 
Total aperture counts	& 682		& 155		& 152		\\ 
Source / total counts (\%)	& 69.9	& 74.0		& 74.4		\\ 
BG-cor. count rate (ks$^{-1}$)	& $4.6 \pm 0.3$ & $1.0 \pm 0.1$	& $1.0 \pm 0.1$	\\ 
\tableline
\end{tabular}
\end{table}

Using XSPEC 12.6.0\footnote{http://heasarc.gsfc.nasa.gov/docs/xanadu/xspec}, we tested 
different spectral models (PL -- \texttt{powerlaw} , BB -- \texttt{bbodyrad} , BB+PL, BB+BB) applying $\chi^2$ statistic. For the photoelectric absorption in the interstellar medium (ISM), we used \texttt{tbabs} with the solar abundance table from \citet{1989GeCoA..53..197A} and the photoelectric cross-section table from \citet{1992ApJ...400..699B} together with a new He cross-section based on \citet{0004-637X-496-2-1044}.
We performed simultaneous fitting of the pn and MOS spectral data. \\

The values of the fits with separate normalizations for each instrument agreed within the 90\% confidence levels with the fit using the same normalization for all instruments. 
For simplicity, we give only the latter in the following . 
The spectral fit parameters were allowed to vary freely for the single component models, while for the combination of the PL with the BB there were not enough counts for constraining all the parameters sufficiently. Therefore we had to freeze a parameter (see below). \\

\begin{figure}[b]
{\includegraphics[height=85mm, bb=70 1 590 740,clip, angle=270]{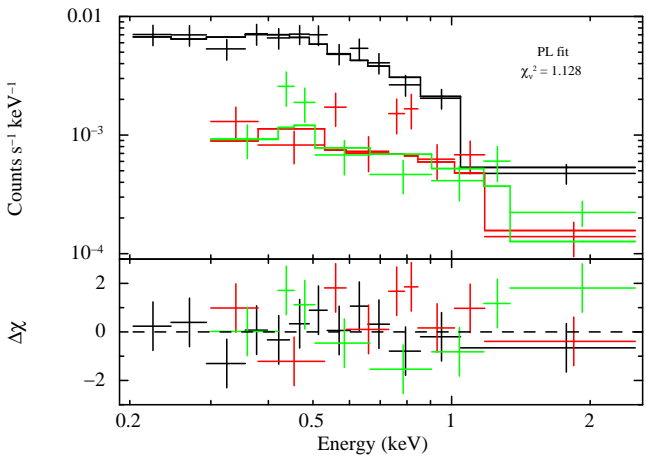}} 
\caption{Absorbed PL fit and its residuals in units of sigmas for the pn (black), MOS1 (red) and MOS2 (green) data.}
\label{plfit}
\end{figure}

\begin{figure}[]
{\includegraphics[width=85mm, bb=40 1 343 244,clip]{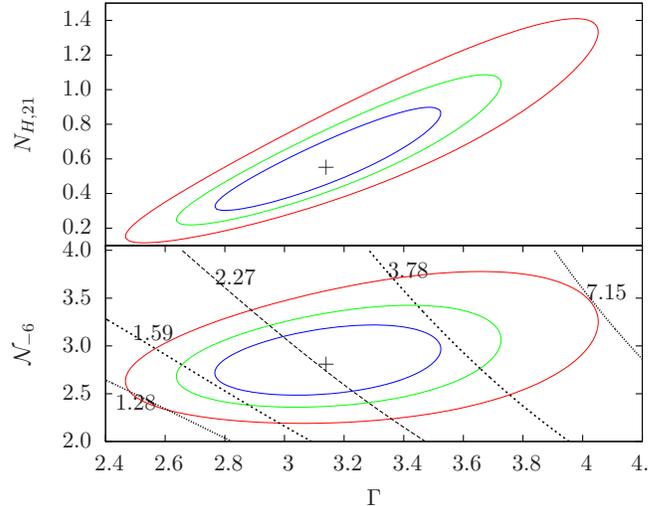}}
\caption{68\%, 90\% \& 99\% confidence contours in the $\Gamma$ -- $N_H$ plane ({\em top}) and in the $\Gamma$ -- PL normalization plane. $N_H$ is units of $10^{21}$\,cm$^{-2}$, the PL normalization ${\cal{N}}_{-6}$ is in units of $10^{-6}$\,photons\,cm$^{-2}$\,s$^{-1}$\,keV$^{-1}$ measured at 1\,keV.
Plotted are also the contours of constant unabsorbed flux, $F^{\rm{unabs}}$ in units of $10^{-14}$ erg cm$^{-2} s^{-1}$ for the 0.2 - 2.5 keV band ({\em bottom}).}
\label{plcont}
\end{figure}

An absorbed PL model with photon index $\Gamma=3.1^{+0.5}_{-0.2}$ and Hydrogen column density  $N_H=(5.5^{+1.6}_{-2.2}) \times 10^{20}$\,cm$^{-2}$ provides a good fit of the data with $\chi_{\nu}^2=1.1$; see Figure~\ref{plfit} for the X-ray spectral fit and Figure~\ref{plcont} for confidence contours for the three model parameters. 
Table~\ref{spectralfits} lists our best spectral fitting parameters.
The confidence levels for each parameter were obtained  using XSPEC's \texttt{error} and \texttt{steppar} commands.
The non-thermal luminosity is estimated to be $L^{\rm{PL}}_{\rm{0.2-2.5\,keV}} =  4 \pi d^2 F^{\rm unabs}_{\rm{0.2-2.5\,keV}} =1.2_{-0.7}^{+1.2}\times10^{29}$ erg s$^{-1}$. 
The luminosity uncertainties include both model and distance uncertainties.\\

\begin{table}[t]
\begin{center}
\caption{Best fit parametric values with 90\% confidence limits. \label{spectralfits}}
\begin{tabular}{lc}
\tableline\tableline
Parameters & Best fit values   \\
\tableline\tableline
Absorbed PL fit &			\\
\tableline
$N_H$ ($10^{20}$ cm$^{-2}$)		& 	$5.5_{-2.2}^{+1.6}$		\\
$\Gamma$			& 	$3.1_{-0.2}^{+0.5}$ 	\\
PL norm ($10^{-6}$\,photons\,cm$^{-2}$\,s$^{-1}$\,keV$^{-1}$ at 1keV) & $2.8_{-0.4}^{+0.4}$	\\
$\chi_{\nu}^2$/d.o.f.				&	1.1/27				\\
$F^{\rm{abs}}_{\rm{0.2 - 2.5\,keV}}$ ($10^{-15}$\,erg\,cm$^{-2}$\,s$^{-1}$) &	$9.4_{-1.0}^{+0.9}$	\\
$F^{\rm{unabs}}_{\rm{0.2 - 2.5\,keV}}$ ($10^{-14}$\,erg\,cm$^{-2}$\,s$^{-1}$) &	$2.3_{-0.8}^{+1.2}$\\	
\tableline
Absorbed PL+BB fit 			&	\\ \tableline
$N_H$ ($10^{20}$ cm$^{-2}$)	&	$2.3_{-2.30}^{+2.4}$ 	\\
$\Gamma$		&	$2.0$ Fixed 			\\
PL norm ($10^{-6}$\,photons\,cm$^{-2}$\,s$^{-1}$\,keV$^{-1}$ at 1keV) &	$1.7_{-0.5}^{+0.4}$	\\
$kT$ (keV)				&	$0.11_{-0.01}^{+0.03}$	\\
$R_{\rm bb}$\tablenotemark{a} (m)	&	$43_{-9}^{+16}$		\\
$\chi^2_\nu$/d.o.f.			&	1.1/26 \\
$F^{\rm{abs}}_{\rm{0.2 - 2.5\,keV}}$ ($10^{-15}$\,erg\,cm$^{-2}$\,s$^{-1}$) &	$9.7_{-1.0}^{+1.0}$	\\
$F^{\rm{unabs}}_{\rm{0.2 - 2.5\,keV}}$ ($10^{-14}$\,erg\,cm$^{-2}$\,s$^{-1}$) &	$1.3_{-0.3}^{+0.5}$\\
\tableline
\end{tabular}
\\
\end{center}
$^{\rm{a}}$ The radius of the blackbody emission was obtained from the normalization using a distance of $d=210$\,pc \citep{Verbiest2012}. Its error is the propagated normalization error only. If one also considers the distance error the correponding values are $R_{\rm bb}= 43_{-14}^{+24}$\,m.
\end{table}

In contrast to the PL fit, a pure thermal model does not describe the data well. 
The best fit of the absorbed BB model has large residuals ($\chi_{\nu}^2 = 2.3$) that rule out the possibility of the emission being entirely BB-like.
In principle, one could try to fit the spectrum with neutron star atmosphere (NSA) models \citep{Zavlin1996,Pavlov1995b}. 
However, the available atmosphere models in XSPEC were calculated either for $B=0$ or strong magnetic fields ($B=10^{12}$\,G and higher). For the magnetic field strength of PSR\,J0108--1431, $B=2.5 \times 10^{11}$\,G, the (redshifted) cyclotron energy is $\sim  2-3$\,keV. This is too close to the observed photon energy range and has a strong effect on the atmophere spectrum, which is not considered in the weak or strong magnetic field models in XSPEC. Therefore, the NSA models are not applicable for PSR\,J0108--1431, and we have to stick to the simplistic BB models.\\

\begin{figure}[]
{\includegraphics[height=85mm, bb=75 11 565 400,clip,angle=270]{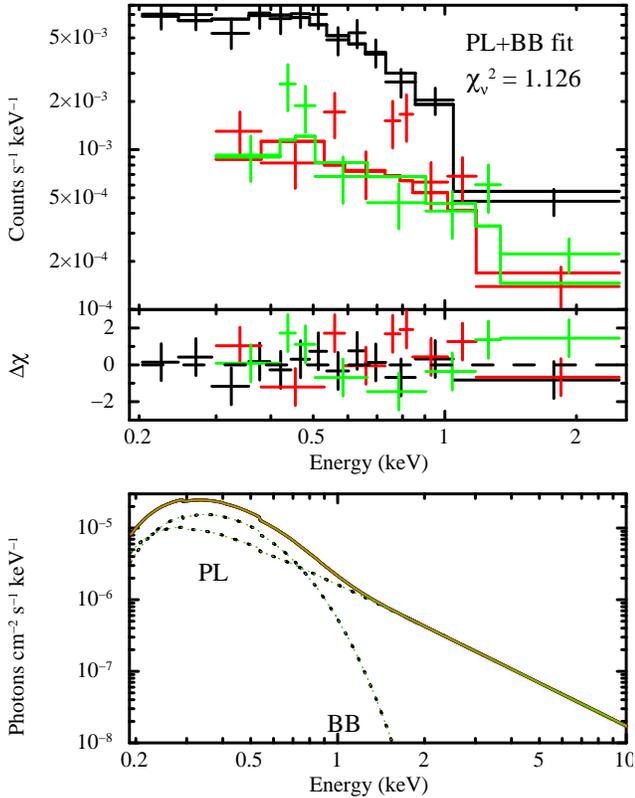}} \\
\caption{{\em Top:} Absorbed PL+BB  fit and the residuals in units of sigma deviations, pn (black), MOS1 (red) and MOS2 (green). {\em Bottom:} Underlying individual PL and BB spectral model components.  
\label{pl2bbrfit}}
\end{figure}

\begin{figure}[]
{\includegraphics[width=85mm, bb=68 22 553 564,clip]{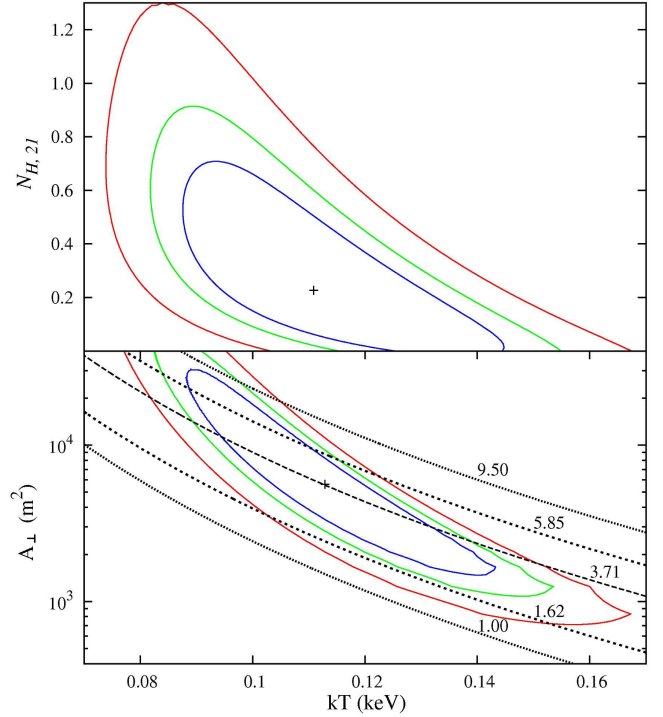}} \\
\caption{68\%, 90\% \& 99\% confidence contours in the $kT$ -- $N_H$ plane ({\em top}) and $kT$ -- $A_{\bot}$ plane ({\em bottom}) for the absorbed PL+BB model. $N_H$ is in units of $10^{21}$ cm$^{-2}$. Lines of constant bolometric luminosities in units of $10^{28}$\,erg\,s$^{-1}$ are overplotted in the $kT$ -- $A_{\bot}$ plane.}
\label{pl2bbrcont}
\end{figure}
 
We also checked a two-component BB+BB model with photoelectric absorption. Such a model could describe the thermal emission from a nonuniformly heated neutron star surface. The best fit for this model also has residuals too large to be acceptable ($\chi_{\nu}^2 = 2.0$).\\

A combination of non-thermal magnetospheric emission and thermal emission from polar caps can be  modeled by a PL+BB model. 
Because of the noise in the data, we had to freeze one fitting parameter.
We froze the photon index to $\Gamma=2$, motivated by the typical photon indices of  younger pulsars, $\Gamma \sim 1 - 2$  \citep{Li2008}. The fit is acceptable ($\chi_{\nu}^2 = 1.1$), its BB temperature is $kT=0.11_{-0.01}^{+0.03}$\,keV.
We use the \texttt{bbodyrad} model in XSPEC, whose normalization factor gives the projected area $A_{\bot}=5700^{+4300}_{-2400}\; d^2_{210}$\,m$^2$ and the corresponding radius of an emitting equivalent sphere, $R_{\rm bb}= (A_{\bot}/\pi)^{1/2} = 43_{-9}^{+16} d_{210}$\,m, where $d_{210}$ is the distance divided by 210\,pc (the errors include only the propagated normalization error), see Table~\ref{spectralfits} for additional fit results.
Since $R_{\rm NS} \gg R_{\rm bb}$, the emission likely comes from part of the surface like hot polar caps.
Figure~\ref{pl2bbrfit} shows the spectral fit and its residuals, while the 68\%, 90\%, and 99\% confidence contours in the $kT$--$N_H$ and $kT$--$A_{\bot}$ planes are plotted in Figure~\ref{pl2bbrcont}. As mentioned above, there are no available atmosphere models for the magnetic field strength of PSR\,J0108--1431. Therefore, we do no fit a two-component PL+NSA model to the data. \\

The luminosity for both components together is estimated to be $L^{\rm{BB+PL}}_{\rm{0.2-2.5\,keV}} = F^{\rm{unabs}}_{\rm{0.2 - 2.5\,keV}} 4 \pi d^2 = 6.8_{-3.7}^{+6.5}\times 10^{28}$ erg s$^{-1}$. The luminosity of the non-thermal component is $L^{\rm{PL}}_{\rm{0.2-2.5\,keV}} = 3.7_{-2.1}^{+3.2}\times 10^{28}$ erg s$^{-1}$. 
The derived values of the temperature and area correspond to the \emph{bolometric}
thermal luminosity of an equivalent sphere $L_{\rm bol}^{\rm bb} = 4 A_{\bot} \sigma T^4 = 3.7_{-2.7}^{+5.8}\times 10^{28}$ erg s$^{-1}$.
The luminosity errors include the propagated distance error and the corresponding propagated fit errors in flux, temperature and emission area.
Lines of constant bolometric luminosities are overplotted in the $kT$ -- $A_{\bot}$ plane (Figure~\ref{pl2bbrcont}, bottom). The above listed errors of $L_{\rm bol}$ encompass nearly the whole $3\sigma$ contours in the  $kT$ -- $A_{\bot}$ plane because we additionally considered the distance error in our conservative error estimate. 

\subsection{Timing Analysis}
\label{timing}
\citet{Hobbs2004} reported the radio ephermerides of PSR\,J0108$-$1431 to be $\nu=1.23829100810(3)$\,Hz and $\dot{\nu}=0.11813(18) \times 10^{-15}$\,Hz\,s$^{-1}$ at MJD\,$50889.0$. At the start time of our X-ray observations, MJD\,$55727.2$, the expected frequency change, $-0.049$\,$\mu$Hz, is negligible for X-ray timing.
The EPIC-pn detector with a nominal frame time of 73.4\,ms in the full frame mode is well suited for the timing analysis of PSR\,J0108$-$1431, while the time resolution of the MOS detectors in  full frame mode, 2.6\,s, does not allow such an analysis. \\

Following up on SAS warnings during the data processing, we checked the EPIC-pn CCD with the target on it for time jumps in the data. 
We set the SAS environment parameter SAS\_JUMP\_TOLERANCE\footnote{SAS\_JUMP\_TOLERANCE is given in units of 20.48\,$\mu$s. Only deviations of the measured actual frame time from the nominal frame time larger than the SAS\_JUMP\_TOLERANCE are identified as time jumps by the SAS.} to a value of 33 to account for the known temperature effect on the actual frame time of the EPIC-pn detector (\citealt{Freyberg2005}, Freyberg et al. 2012\footnote{See also: http://www2.le.ac.uk/departments/physics/research/\protect\\src/Missions/xmm-newton/technical/leicester-2012-03/freyberg-cal-2012.pdf}
). 
In addition, we excluded the end of the observation ($>119.28$\,ks after start). At that time, a bright X-ray background flare caused the instrument to switch to the counting mode\footnote{XMM-Newton Users Handbook, section 3.3.2 -- Science modes of the EPIC cameras} which resulted in finding of false time jumps by the SAS.
Our final data for the timing analysis were free of apparent time jumps.
All times-of-arrival of the X-ray photons were corrected to the solar barycentric system using the standard task \texttt{barycen}.

We used the $Z_n^2$ test, the sum of powers of first $n$ harmonics, (e.g., \citealt{Buccheri1983}) to search for pulsations in these data. For our timing analysis, we checked 7 different GTI screenings, 11 different energy regions, and 5 different extraction regions in order to maximize the $Z^2_1$.
The GTI screening with pn count rate $< 4$ counts\,s$^{-1}$ for a 100\,s binned background light curve (10--12\,keV), the extraction radius of 8\,$^{\prime\prime}$, and the energy range of 0.15--2\,keV were found to be the optimal choices among the tested variants. At the chosen GTI filter, there is only one small (200\,s) time gap in the last fifth of the exposure; see Figure~\ref{PNrate}.
This filtering provided 507 events during a time span of $T_{\rm span}=119.1$\,ks.\\ 

PSR\,J0108$-$1431 has been monitored for 16 years, and no glitches have been detected \citep{Espinoza2011}.
Investigating possible timing irregularities for 366 pulsars, \citet{Hobbs2010} found PSR\,J0108$-$1431 to be a very stable pulsar with low timing noise like other pulsars with similarly low $\dot{\nu}$. Therefore, we could search for X-ray pulsations at the radio pulsation frequency.
However, to check whether the above-mentioned time jump correction worked properly,
we searched a frequency range of $1.2382$--$1.2384$\,Hz with a sampling of 0.1\,$\mu$Hz, thus oversampling the expected $Z_n^2$ peak width, $\sim T^{-1}_{\rm span}=8.4$\,$\mu$Hz,  by a factor of 80.
A peak $Z^2_{1,\rm{max}} = 48.0$  is found at $\nu_{0}=1.2382908$\,Hz $\pm 0.7$\,$\mu$Hz, corresponding to a period of 0.8075647\,s $\pm 0.5$\,$\mu$s.
The frequency uncertainty was derived similarly to \citet{Chang2012} as $\delta \nu=0.55 T^{-1}_{\rm span} (Z^2_{1,\rm{max}})^{-1/2}$.
Within errors the X-ray pulse frequency agrees with the radio pulse frequency very well. 
The probability to find $Z^2_{1,\rm{max}} = 48.0$ by chance is $p= \exp(-Z^2_{1,\rm{max}}/2)=4\times 10^{-11}$, which corresponds to the $6.6\sigma$ confidence level.\\
\begin{figure}[t]
\begin{center}
\includegraphics[width=8.5cm,bb=25 7 835 400,clip]{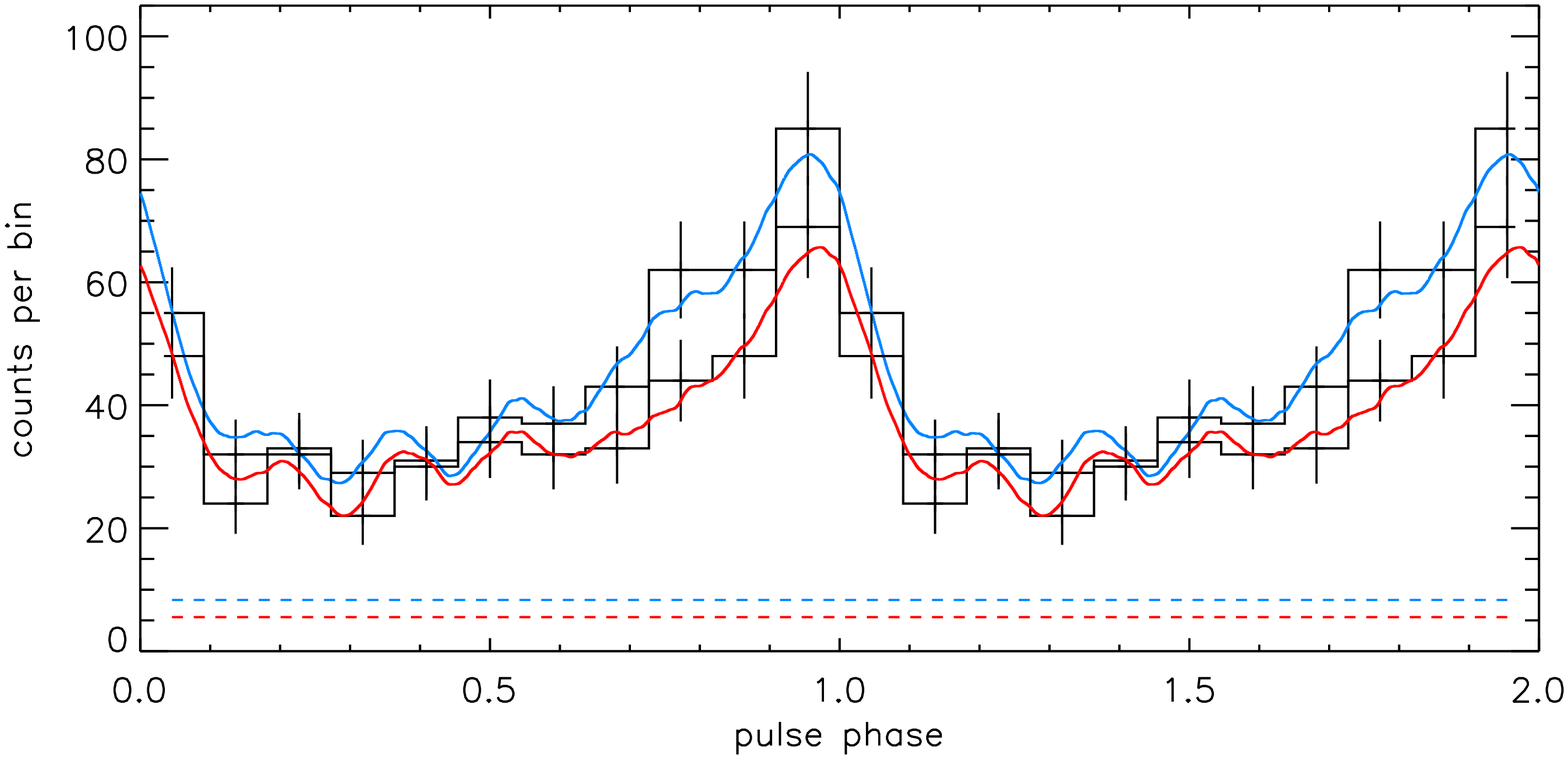}
\includegraphics[width=8.5cm,bb=25 7 835 400,clip]{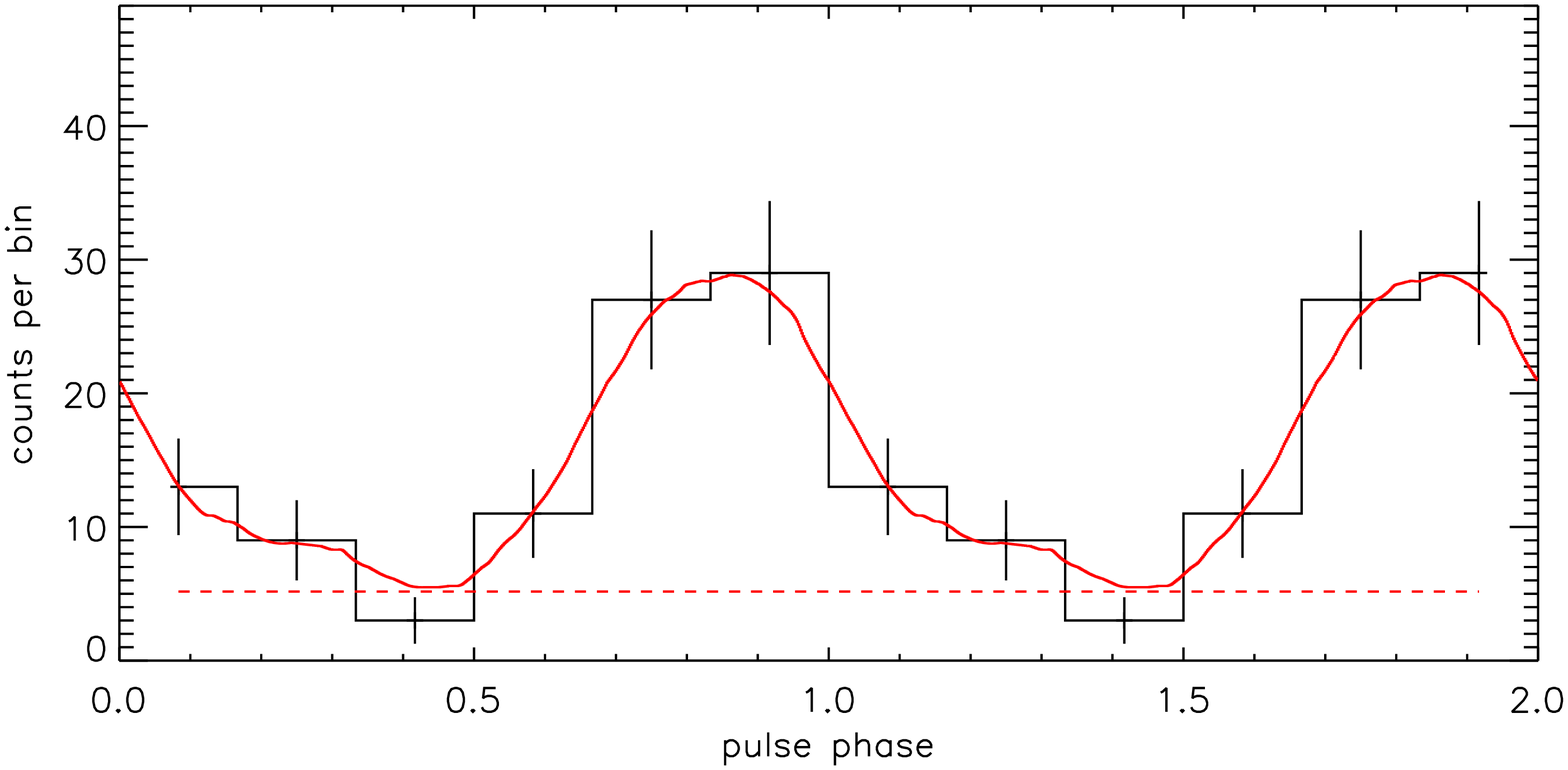}
\caption{Folded light curves as histograms and reference phase averaged (RPA) pulse profiles  (see text). For a better overview, two full phase cycles are shown.\protect\\ {\bf Upper panel: }
The upper 11-bin histogram represents the full considered energy range, 0.15--2\,keV, the corresponding RPA pulse profile is shown in blue, its average background level is indicated with a blue dashed line. Similarly, for the energy range 0.15--1\,keV, the red RPA  pulse profile  corresponds to the lower histogram, and the average background level for this energy range is indicated with a red dashed line. {\bf Lower panel: }The 6-bin histogram and  the RPA  pulse profile represent the folded light curve in the energy range 1--2\,keV.}
\label{foldedlc1}
\end{center}
\end{figure}

To explore the harmonic content of the X-ray pulsations and refine the significance estimate, we performed the $Z_n^2$ test accounting for up to $n = 7$ harmonics. 
The peak $Z_n^2$ values (and the respective significances) are 
$Z^2_2 = 57.9$ ($6.8\sigma$),
$Z^2_3 = 61.5$ ($6.7\sigma$),
$Z^2_4 = 66.4$ ($6.7\sigma$),
$Z^2_5 = 75.3$ ($6.9\sigma$),
$Z^2_6 = 75.5$ ($6.6\sigma$),
$Z^2_7 = 80.0$  ($6.4\sigma$) at frequencies consistent within errors with the one found from the $Z^2_1$ test. 
For the application of the H-test \citep{Jager1989}, $H={\rm max} \{H_m \}$, we found the $H_m=Z^2_m-4m+4$ to be  48.0, 53.9, 53.5, 54.4, 59.3, 55.5, 56.0 for the first to seventh harmonic, respectively, i.e. the pulsation have statistically significant contributions from 5 leading harmonics of the principal frequency.\footnote{We checked $H_m$ up to 20 harmonics as recommended by \citet{Jager1989} with the same result.}
Thus, the pulsations of PSR\,J0108$-$1431 are unambiguously detected in X-rays with a siginificance of $6.9\sigma$.\\

In order to visualize the pulse shape, we obtained a folded light curve in form of a histogram for the energy range
0.15--2\,keV (507 counts), as well as for the splitted energy ranges 0.15--1\,keV (416 counts) and 1--2\,keV (92 counts). 
The frequency $\nu_{0} =1.2382908$\,Hz, was used for the folding. 
We applied the so-called Scott's rule for setting the \emph{upper} bound for the histogram bin width by \citet{Terrell1985}. These authors concluded that their rule to avoid oversmoothing gives nearly optimal results for a variety of smooth probability densities. However, choosing the optimal histogram bin size remains a matter of debate in statistical data analysis (for a review, see e.g., \citealt{Scott1992}). According to Scott's rule the number of bins must be $ \gtrsim (2N)^{1/3}$, where $N$ is the number of events. We selected $N_{\rm hist}=11$ bins for the energy ranges 0.15--2\,keV and 0.15--1\,keV, and $N_{\rm hist}=6$ for the energy range 1--2\,keV.\\

The number of counts in the histogram bins depends on the reference phase, which was chosen arbitrarily. The histogram can look quite different for another choice of reference phase, especially if there are few bins. 
To obtain the folded light curve independent of the reference phase choice, we averaged the histogram over the reference phase\footnote{A similar phase averaging was applied by \citet{Zavlin2002s} (their Figure 5).}, varying the latter within one histogram bin; see the Appendix~\ref{smooth} for a short description of the used algorithm.
In the following, we call the obtained new histogram the reference phase averaged (RPA) pulse profile.
The histograms of the folded light curve together with the respective  RPA pulse profiles are shown in Figure~\ref{foldedlc1}. The pulse shape for the full energy band, 0.15--2\,keV, and the soft energy band appear to be asymmetric, with a slower rise and steeper decay. The two RPA pulse profiles have their maxima at similar phases. The 1--2\,keV RPA pulse profile appears more sinusodial and slightly shifted to smaller phases compared to the low-energy and broad bands. 
However, the pulse shape in the energy range 1--2\,keV has poorer statistics, and direct comparisons must be regarded with caution.
Using the  RPA pulse profiles as probability distribution templates and the respective number of events as sample sizes, we carried out Monte Carlo simulations of the profiles to estimate the uncertainties of the phase positions of the respective RPA pulse profile maxima.
The maxima of the  RPA pulse profiles in Figure~\ref{foldedlc1} are at $\phi=0.96 \pm 0.03$, $0.97 \pm 0.03$, and $0.86 \pm 0.06$ for the energy ranges 0.15--2\,keV, 0.15--1\,keV, and 1--2\,keV, respectively (see Figure~\ref{foldedlc1}).\\

\begin{figure}[]
\begin{center}
\includegraphics[width=8.5cm,bb=37 24 565 280,clip]{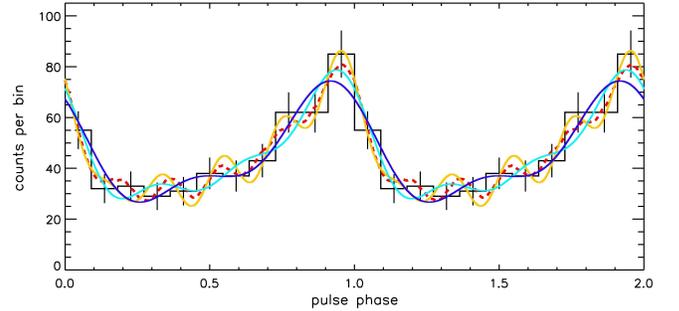}
\caption{Folded light curve histogram for the full considered energy range 0.15--2\,keV together with the Fourier series including up to three harmonics (dark blue), four harmonics (cyan) and five harmonics (yellow).   
For comparison, the  RPA pulse profile from Figure~\ref{foldedlc1} is also plotted as red dashed curve.}
\label{foldedlcfs}
\end{center}
\end{figure}

For the full energy band, 0.15--2\,keV, for which we have the best statistics, we also use  another  visualization method.
We apply Fourier decomposition $F_m(\phi)$ up to an harmonic $m$ to describe the pulse shape. 
\begin{equation}
 F_m(\phi)=\frac{N}{N_{\rm hist}} \left[ 1+2 \sum^m_{k=1} a_k \cos (2 \pi k \phi) + b_k \sin (2 \pi k \phi) \right]
\end{equation}
where
\begin{eqnarray}
a_k = \frac{1}{N} \sum^N_{i=1} \cos (2 \pi k \nu t_i)  {\rm , } \hspace{0.2cm} 
b_k = \frac{1}{N} \sum^N_{i=1} \sin (2 \pi k \nu t_i)  \hspace{0.4cm} 
\end{eqnarray}
are one half of the empirical Fourier coefficients, $\phi$ is the phase, $N$ is the total number of events in the folded lightcurve, $t_i$ are the event times, and $\nu=\nu_{0}=1.2382908$\,Hz is the pulsation frequency.
The scaling factor $N/N_ {\rm hist}$ allows one to compare the Fourier series plot with the $N_{\rm hist}$-bin histogram. Here, we apply a scaling factor $N/N_ {\rm hist}=507/11 \approx 46$ to
 plot the Fourier series for $m=$3, 4 and 5 in Figure~\ref{foldedlcfs}. 
While including five harmonics introduces `wiggles' of the same order as the errors in a light curve bin, Fourier series with coefficients including up to three or four harmonics appear to model well the folded light curve, in particular its asymmetric pulse shape.\\

From the histograms, we calculate the empirical pulsed fractions as the ratio of the number of counts above the minimum to the total number of counts. For the energy ranges 0.15--2\,keV, 0.15--1\,keV, and 1-2\,keV the pulsed fractions are $f_p=39$\,\%$ \pm 6$\,\%, $33$\,\%$\pm 7$\,\%, and $80$\,\%$ \pm 15$\,\%, and the background-corrected, instrinsic pulsed fractions are $f_p^{int}=47$\,\%$ \pm 7$\,\%, $38$\,\%$ \pm 8$\,\%, and $100^{+0}_{-18} $\,\% respectively.
We calculated the errors of the pulsed fraction as $\delta f_p = (2/N)^{\frac{1}{2}}$, following the estimate for broad pulses by \citet{Jager1994}\footnote{As the power in the fundamental harmonic strongly exceeds the powers in higher harmonics for PSR\,J0108$-$1431, this uncertainty estimate is applicable, albeit approximate.}.

\section{Complementary radio data} 
It is interesting to compare the radio pulse with the X-ray pulse, in
particular the phase difference between them. 
We use the monitoring radio data described by \citet{Weltevrede2010} for our comparison of the radio and the X-ray profiles. Observations at 1.4\,GHz were
obtained from December 2009 to July 2012 using the 64-m radio
telescope at Parkes, NSW, Australia. The data were calibrated and times-of-arrival (TOAs)
produced using \texttt{psrchive} routines \citep{Hotan2004} and the
\texttt{tempo2} package \citep{Hobbs2006}. An
accurate value of the radio dispersion measure (DM) is necessary to
correct for the dispersion delay when comparing the relative
phases of the 1.4\,GHz radio and high-energy pulse profiles. The
measured DM for PSR\,J0108$-$1431 is $2.38 \pm
0.19$\,cm$^{-3}$\,pc \citep{Hobbs2004}. This translates into an
DM-induced uncertainty of $\sim 402$\,$\mu$s or $5 \times 10^{-4}$ in
the phase of the radio pulse.

\begin{figure}[] 
\begin{center} 
\includegraphics[width=9cm,bb=30 12 730 402,clip]{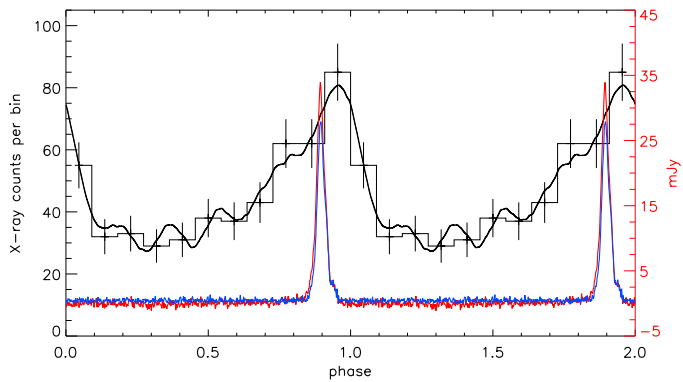} 
\caption{The folded X-ray light curve histogram (energy range 0.15--2\,keV) is plotted together with the corresponding  RPA pulse profile (both in black). The red curve shows the total intensity of the 1.4\,GHz radio pulse in mJy; the blue curve shows the linearly polarized part of the radio pulse. The reference phases of the X-ray and radio profiles were aligned as described in the text.} \label{radioXray} 
\end{center} 
\end{figure}

For the comparison with the radio pulse shape, we converted the
TOAs of the X-ray photons to the solar system
barycenter using the standard SAS task \texttt{barycen} with the same
coordinates as in the radio \texttt{tempo2} analysis and the same
DE405 solar-system ephemeris \citep{Standish1998}. 
We used the barycentric TOAs of five X -ray events in \texttt{tempo2} to derive
their phases relative to the radio profile. Knowing the X-ray phases
of these events as well, we determined the offset between the radio
and X-ray phase reference points to be $0.898 \pm
0.003$. Correcting for this shift between the reference systems, Figure~\ref{radioXray} shows the X-ray and radio profiles together in the same phase range. 
We confirm the unusual
high linear polarization of the radio pulse reported by
\citet{Weltevrede2010}, obtaining a value of 77\% for the fractional
linear polarization at the pulse peak. 
Using the peak positions of the RPA pulse profiles from Section~\ref{timing}, the radio pulse leads the X-ray pulse peak by
$0.06 \pm 0.03$ and $0.08 \pm 0.03$ in the energy ranges 0.15--2\,keV and 0.15--1\,keV, respectively. In contrast, the X-ray pulse at 1--2\,keV leads the radio pulse by $0.03 \pm 0.06$. 
The quoted errors take into account the
uncertainties in the reference phase shift, the DM-induced phase
uncertainty, the absolute XMM-$Newton$ EPIC-pn timing accuracy of
$48$\,$\mu$s \citep{Martin2012}, and the dominating error of the maximum position in the RPA pulse profile (Section~\ref{timing}). 

\section{Discussion}
\subsection{The X-ray spectrum and luminosity}
\label{discussspectrum}
PSR\,J0108$-$1431 is the oldest among the non-recycled ordinary pulsars detected in X-rays, with the spin-down age a factor of 4 larger 
than that of PSR\,B1451$-$68, the previous record holder. 
It also has the lowest $\dot{E}$ among the same sample of X-ray detected non-recycled pulsars. Our  new {\sl XMM-Newton} observation provided a factor of 20 more source counts enabling better characterization of the pulsar's spectrum compared to the previous {\sl Chandra} observation. 
However, the strong flaring and high background at energies above $\sim 2$ keV have undermined energy-resolved timing
and phase-resolved spectral analysis of the pulsar emission.\\ 

Formally, a simple absorbed PL model 
and an absorbed PL+BB model describe the data equally well.
The photon index $\Gamma=3.1^{+0.5}_{-0.2}$ (see also Figure~\ref{plcont}) of the PL fit is larger (i.e., the spectrum is steeper) than 
$\Gamma \sim 1$--2, typical for young pulsars \citep{Li2008}. As mentioned above, for the BB+PL fit we had to freeze one parameter 
because of the small number of counts and strong noise.
 We chose to fix the photon index at $\Gamma=2$, assuming similar magnetospheric emission characteristics for young and old pulsars.\\

The PL fit falls in line with the trend observed in other old pulsars,  for which the PL fits suggest 
too large $N_H$ values in conjunction with 
larger photon indices $\Gamma$. 
For the line of sight of PSR J0108--1431, the LAB Survey of Galactic neutral hydrogen reports $N_{H\sc{I}}=2.1\times 10^{20}$\,cm$^{-2}$ in this direction \citep{Kalberla2005}, the \citet{Dickey1990} neutral hydrogen survey reports $N_{H\sc{I}}=1.8 \times 10^{20}$\,cm$^{-2}$.
The $N_H =  5.5_{-2.2}^{+1.6} \times 10^{20} $\,cm$^{-2}$ from our PL fit is above these total Galactic values.
We use the Lutz-Kelker corrected, parallactic distance $d = 210^{+90}_{-50}$ pc by \citet{Verbiest2012} 
to estimate the expected $N_H$ value applying the `analytical' 3D extinction model described by \citet{Posselt2007}.  For the close ($<250$\,pc) solar neighbourhood, this model is based on the 3D Na\,D absorption line mapping by \citet{Lallement2003} and has a resolution of $\sim 25$\,pc; at larger distances an analytical model is used for the extinction (see  \citealt{Posselt2007,Posselt2008} for more details).
Considering the errors of the distance, we derived an expected $N_H$ is in the range 
(0.3--$0.8) \times 10^{20} $\,cm$^{-2}$.
Assuming 10 H atoms per electron, we derive a similarly low expected $N_H^{\rm DM}=0.7 \times 10^{20}$\,cm$^{-2}$ from the pulsar dispersion measure, ${\rm DM}=2.38$\,pc\,cm$^{-3}$ \citep{Hobbs2004}.
Thus, 
the absorption is overestimated in the simple PL fit
(see Figure 5) while the 90\% confidence range of the $N_H$ parameter in the BB+PL fit includes the expected $N_H$ value.\\

The temperature obtained in the BB+PL fit, $T=1.3$\,MK, is a reasonable estimate for the expected heated polar cap region. 
The effective projected emitting area is $A_{\bot}=5700$\,m$^2$. This is a factor of $\sim 14$ smaller than the conventional polar cap area $A_{\rm pc} = 2\pi^2R_{\rm NS}^3/(cP) \approx 82000$\,m$^2$, using  $R_{\rm NS}=10$\,km.
PSR\,J0108$-$1431 is very similar to other old pulsars in this respect \citep{Posselt2012, Pavlov2009, Misanovic2008, Kargaltsev2006a, Zhang2005}. 
We should note, however,
that the values of the temperature and, particularly, the projected area are rather uncertain because of their strong correlation (Figure~\ref{pl2bbrcont}).\\

As mentioned in the Introduction, one has to take into account that thermal emission from the surface layers of a
 neutron star can differ substantially from the BB emission (e.g., \citealt{Pavlov1995b}). 
 In particular, if the emission emerges from a light-element 
(H or He) atmosphere, fitting with neutron star atmosphere models may yield
the effective temperature a factor of 2 lower than $T_{bb}$, 
and the projected emitting area a factor of 10--100 larger than $A_{\bot}$, while the bolometric luminosity does not change substantially. 
The magnetic field can have a strong impact on the observed emission of a neutron star atmosphere. The current neutron star atmosphere models available in XSPEC are not applicable to PSR J0108--1431 because the electron cyclotron line, caused by its magnetic field, $B=2.5 \times 10^{11}$\,G, is in the observable energy range, but it is not included in the XSPEC models. Therefore, we do not discuss atmosphere models for PSR J0108--1431.\\

Overall, there are now several old pulsars 
whose spectra could be described by the PL model, 
but the properties of these fits  -- in particular, the 
large $\Gamma$ and the much higher than  expected $N_H$ -- 
suggest a more complicated model.
Adding a thermal (e.g., BB) component to the spectral model allows lower $N_H$ values and, correspondingly, smaller $\Gamma$, similar to those of
younger pulsars.\\

\begin{figure}[]
\includegraphics[width=6.5cm,bb=66 80 530 695,clip, angle=90]{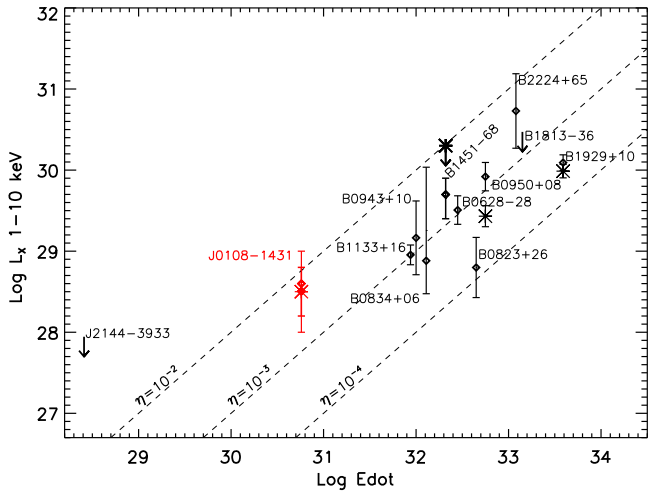}
\caption{X-ray luminosities and upper limits of eleven old pulsars versus their spin-down energies $\dot{E}$. 
In cases of sufficient numbers of counts, the diamonds and asterisks show the 
1--10 keV non-thermal luminosities and bolometric thermal luminosities, respectively; otherwise the luminosities were obtained from PL fits. The arrows mark upper limits derived from X-ray non-detections. 
 This figure is an update of the Figure~4 presented by \citet{Posselt2012} using  distances corrected for the Lutz-Kelker bias by \citet{Verbiest2012}.
In case of PSR\,B1451--68, error propagation leads to a lower error of the bolometric thermal luminosity as large as the best-fit value. Therefore, we plot only an upper limit for the bolometric thermal luminosity of PSR\,B1451--68.}
\label{oldlum}
\end{figure}

For comparison with other pulsars, we calculated the non-thermal (PL) luminosity in the 1--10 keV energy range:
$L^{\rm PL}_{\rm{1-10\,keV}} = 2.0_{-1.1}^{+1.8} \times 10^{28} $\,erg\,s$^{-1}$ for the PL fit,
 and  $L^{\rm PL}_{\rm{1-10\,keV}} = 3.3_{-1.9}^{+3.0} \times 10^{28} $\,erg\,s$^{-1}$ for the PL component in the BB+PL fit\footnote{The  1--10 keV
PL component luminosity and its errors were estimated for the fixed $\Gamma=2$.}. 
The higher photon index of the PL fit, obtained from the fitting at a lower energy range, is the reason why the $L^{\rm PL}$ is smaller in the case of the PL fit than in the case of the PL+BB fit.  
The errors were calculated from the 90\% confidence errors of the unabsorbed flux and the distance uncertainty of the Lutz-Kelker corrected distance as listed by \citet{Verbiest2012}.
These luminosities translate into non-thermal  X-ray efficiencies of  $\eta^{\rm PL}_{\rm 1-10\,keV}= L^{\rm PL}_{\rm 1-10\,keV}/\dot{E} \sim 0.003 \, d_{210}^2$  and  $\sim 0.006 \, d_{210}^2$ for the PL fit and the PL+BB fit, respectively.
Using the bolometric luminosity $L_{\rm bol}^{\rm bb} = 3.7^{+5.8}_{-2.7}\times 10^{28}$ erg s$^{-1}$ (Section~\ref{spectralana}), one can also estimate the thermal polar cap heating efficiency from the PL+BB fit,
$\eta^{\rm bb}_{\rm PC} = L^{\rm bb}_{\rm bol}/\dot{E} \sim 0.006 \, d_{210}^2$, 
which gives the fraction of spindown power heating the polar caps.
\citet{Harding2001} predicted that the expected polar cap heating efficiency for ordinary pulsars for ages $\tau \lesssim 10^7$\,years grows with age and period (e.g., $\eta_{PC}\sim 0.01$ for $\tau=1.5 \times 10^7$\,years, $P=0.2$\,s; see their Figure\,7). For pulsars with $\tau > 10^7$\,years they cautioned that the pulsar cannot produce enough electron-positron pairs to fully screen the electric field. Thus, the fraction of the returning positrons that heat the polar cap decreases, and the heating efficiency will drop.  
Comparing the estimated polar cap heating efficiency with the predictions by \citet{Harding2001}, PSR\,J0108$-$1431 has a lower efficiency than one would expect for a pulsar having the same period (0.8\,s) at an age of $\tau = 10^7$\,years.
This could indicate that the returning positron fraction is indeed smaller for PSR\,J0108$-$1431 than for younger pulsars.\\

In Figure~\ref{oldlum}, we updated the X-ray luminosities versus spin-down power plot by \citet{Posselt2012} (their Figure\,4\footnote{Note that \citet{Posselt2012} used distances derived from the Lutz-Kelker corrected parallaxes \citep{Verbiest2010}, which differ from the Lutz-Kelker corrected distances. See \citet{Verbiest2012} for details about the differences.}). The pulsar again
 shows properties comparable to those of other old pulsars. 
In particular, its non-thermal X-ray efficiency is higher than those of young pulsars, most of which have $\eta^{\rm PL}_{\rm 1-10\,keV} < 10^{-3}$.
Thus, the observation that older pulsars seem to radiate in X-rays more efficiently than younger ones \citep{Zharikov2004, Kargaltsev2006a} is reinforced.

\subsection{The X-ray pulsations of PSR J0108--1431}
The timing analysis of \psrj \,establishes, for the first time, unambiguous X-ray pulsations from this pulsar. More than 2 harmonics are required to explain the asymmetric pulse profile.
Such asymmetric pulse profiles can have several explanations, different for
nonthermal and thermal emission.\\ 

Asymmetric pulse profiles can be easily produced by nonthermal emission
in the outer gap model \citep{Romani1995}, especially if the
special relativity effects, such as aberration and retardation, are taken
into account 
(e.g., \citealt{Dyks2003} and references therein).
A comparison of the detected pulse shapes with those predicted by the
models of magnetospheric high-energy emission could be useful in establishing
the geometry of the emitting region, especially if additional information about
directions of the pulsar's  magnetic and spin axes is available from, e.g.,
radio-polarimetry. 
For PSR\,J0108$-$1431, we do not know if the bulk of the observed X-ray emission is produced in the pulsar's magnetosphere (as it is implied by
the PL spectral model), or whether only a high-energy tail originates from
the magnetosphere (as we assumed in the PL+BB model).
A narrow, sharp pulse would unambiguously point to strongly beamed magnetospheric emission; however, the pulse profile appears to be rather broad (Figure~\ref{foldedlc1}) even at higher energies (1--2\,keV).  
Thus, we can neither rule out nor confirm purely non-thermal emission as the cause for the asymmetric pulse profile.

The asymmetric pulse profile can also be produced by thermal emission if
the distributions of temperature and/or magnetic field over the neutron star 
surface are not axially symmetric. For instance, an asymmetric pulse with a shape similar to the observed one could be produced by a polar cap whose trailing edge is hotter (hence brighter) than the leading edge, as it was suggested by \citet{McGowan2003} to explain a similar pulse shape for the young ($\tau = 30$ kyr) PSR J0538+2817.
Another possibility to explain the asymmetric pulse of thermal radiation is
provided by the anisotropy in the local intensity of radiation emerging from 
a neutron star atmosphere. 
In contrast to the BB radiation, atmospheric 
radiation in a strong magnetic field is highly anisotropic, with a `pencil component' along the magnetic field, and `fan component' transverse to the magnetic field \citep{Pavlov1994}. 
If the magnetic field is not perpendicular to the
polar cap surface (which can occur if the field is essentially nondipolar), then a great variety of pulse shapes can be produced, depending on the magnetic field inclination and the angle between the spin axis and line of sight.
The strong anisotropy of the atmospheric emission could also explain the relatively
high pulsed fraction of the thermal emission from PSR\,J0108$-$1431. The pulsed fraction $f^{int}_p=38$\,\%$ \pm 8$\,\%  in the energy range 0.15--1\,keV appears to be too high for locally isotropic 
BB emission, whose pulsations would be strongly smeared by the bending
of photon trajectories in the gravitational field of the neutron star \citep{Zavlin1995a}.
Thus, the properties of the pulsations detected in PSR\,J0108--1431 do not
contradict to the presence of a thermal component in its emission. 
To firmly prove the presence of this component and infer the properties of the polar cap(s) from the comparison with the models, a refined energy-resolved timing
analysis is needed, which would require ``cleaner'' data.\\

We found a possible phase shift $\Delta\phi = 0.06\pm 0.03$ between  the 1.4\,GHz and 0.15--2\,keV pulse peaks. In principle, it might be due to different 
emission heights, but the uncertainty of $\Delta\phi$ is too large to warrant 
such an investigation.
It is interesting to note, however, that the 1--2\,keV RPA pulse profile peak appears to be  closer to the radio peak than the 0.15--1\,keV RPA pulse profile peak. 
The 1--2\,keV pulse seems to be slightly shifted by $\Delta\phi \sim 0.1$ from the 0.15--1\,keV pulse (considering the maxima of the RPA pulse profiles).
In addition, as is apparent from Figure~\ref{foldedlc1}, the pulse profile at 1--2\,keV appears to be sinusodial while  the soft energy pulse profile (0.15--1\,keV) is asymmetric.  
Phase shifts and different pulse shapes would support the hypothesis of two components
-- an (anisotropic) thermal component and a non-thermal component.
However, the high background and the 
associated large errors prohibit any firm conclusion. 
It remains to be confirmed, whether or not the two-component model is a viable interpretation,
in particular, if a separate non-thermal X-ray component 
is formed in close vicinity to the radio emission.

\section{Summary}
We detected for the first time X-ray pulsations of PSR\,J0108$-$1431 at the pulse frequency expected from radio pulse timing.
The pulse shape is rather asymmetric, requiring up to 5 harmonics to describe it. The peaks of the radio and the X-ray pulses are close to each other in phase.
The X-ray spectrum and the high non-thermal X-ray efficiency of PSR\,J0108$-$1431 are comparable to other old pulsars. In particular, while the spectrum can formally be well described by a PL fit, the expected smaller photon index and lower $N_H$ are better accounted for in a PL+BB model.
We were not able to investigate phase-resolved spectra, because of the high background in the strongly flare-impaired observation.

\acknowledgments
We thank M. Freyberg and F. Haberl for their very valuable explanations regarding XMM$-Newton$ timing and timing jumps found by the SAS.
We also thank M. Kerr for discussions about the $H$-test.\\
This work was partly supported by NASA grants NNX12AE09G and NNX09AC84G, and by the Ministry of Education and Science of the Russian
Federation (contract 11.G34.310001).\\
Based on observations obtained with XMM-Newton, an ESA science mission
with instruments and contributions directly funded by
ESA Member States and NASA.
The Parkes radio telescope is part of the Australia Telescope which is
funded by the Commonwealth Government for operation as a National Facility
managed by CSIRO. This research has made use of SAOImage DS9, developed by
SAO; the SIMBAD and VizieR databases, operated at CDS, Strasbourg, France;
and SAO/NASA's Astrophysics Data System Bibliographic Services.

\appendix
\section{Obtaining the reference phase averaged pulse profile}
\label{smooth}

In the following, we briefly describe the sliding average bin algorithm used to average over  reference phases within one histogram bin, producing the reference phase averaged (RPA) pulse profile.
First, we create $k=0, 1, \ldots, (N_{\rm sub}-1)$ folded light curve histograms with
$N_{\rm hist}$ bins by shifting the reference time in the time series by
${\Delta t}_k = kP/(N_{\rm hist} \times N_{\rm sub})$, where $P$ is the period of the pulsations, and $N_{\rm sub}>1$ is an integer, which is equivalent to shifting the reference phase by $\Delta\phi_k = \Delta t_k /P$.
As usual, the bin heights of the histograms are the number of photon counts in the respective bins. 
Second, we divide the whole phase interval into $n=1,2, \ldots N_{\rm sub}\times N_{\rm hist}$ sub-intervals with bin heights $h^k_n$, i.e., there are $N_{\rm sub}$ sub-intervals in  each original bin having all the same bin heights.
Finally, we average the bin heights of the $N_{\rm sub}$ histograms in each of the sub-intervals:
\begin{equation}
 h^{\rm RPA}_{n}= \frac{1}{N_{\rm sub}} \sum^{N_{\rm sub}-1}_{k=0} h^k_n
\end{equation}
For our histograms and RPA pulse profiles in Figure~\ref{foldedlc1}, we used $N_{\rm sub}=100$, and, as mentioned in Section~\ref{timing}, we used $N_{\rm hist}=11$ bins for the energy ranges 0.15 to 2\,keV and 0.15--1\,keV, and $N_{\rm hist}=6$ for the energy range 1-2\,keV.\\

\bibliographystyle{apj}
\bibliography{generalI}

\end{document}